\documentclass[runningheads]{llncs}
\usepackage[T1]{fontenc}
\usepackage{graphicx}
\usepackage{xcolor}
\usepackage{amssymb}
\usepackage{pifont}
\usepackage{booktabs}
\usepackage{tikz}
\usepackage{subcaption}
\usepackage{url}
\usetikzlibrary{patterns}
\usepackage{multirow}

\usepackage[most]{tcolorbox}
\usepackage{tabularx}
\usepackage{multirow}
\usepackage{array}

\usepackage{booktabs}
\usepackage{dblfloatfix}   % helps table* placement in two-column mode
\usepackage{array}
\usepackage{makecell}

\usepackage{float}
\usepackage{placeins}

\definecolor{symgreen}{RGB}{46, 125, 50}   % 深绿色
\definecolor{symviolet}{RGB}{123, 31, 162} % 紫色
\definecolor{symred}{RGB}{198, 40, 40}     % 红色

\newcommand{\symP}{
\tikz{\fill[symgreen] (0,0) rectangle (0.18,0.18);}
}
\newcommand{\symS}{
\tikz{
\draw[symviolet, line width=0.4pt] (0,0) rectangle (0.18,0.18);
\fill[symviolet, pattern=north east lines, pattern color=symviolet]
(0,0) rectangle (0.18,0.18);
}
}
\newcommand{\symN}{
\tikz{\draw[symred, line width=0.5pt] (0,0) rectangle (0.18,0.18);}
}
\begin{document}

\title{SoK: Blockchain Agent-to-Agent Payments}

%\orcidID{0009-0007-1149-6098} 
%\orcidID{0000-0001-5959-4817}
%\orcidID{0009-0005-4610-6550}
%\orcidID{0000-0003-2264-749X}
%\orcidID{0000-0002-0629-6792}
%\orcidID{0009-0001-0197-1875}
%\orcidID{0000-0001-8006-7392}

% Spiridon Zarkov,
% Zark Lab, Singapore
% Email: spiridon@zarklab.ai

\author{Yuanzhe Zhang\inst{1}\and
Yuexin Xiang\inst{2} \and
Yuchen Lei\inst{3} \and Qin Wang\inst{4} 
\and \\ Tian Qiu\inst{1}\and Yujing Sun\inst{1}\and Spiridon Zarkov \inst{5} 
\and Tsz Hon Yuen\inst{2}\and \\ Andreas Deppeler\inst{6}\and Jiangshan Yu\inst{7}
\and Kwok-Yan Lam\inst{1}
}
\authorrunning{Zhang et al.}
\institute{Digital Trust Center, Nanyang Technological University, Singapore 
\and
Faculty of Information Technology, Monash University, Australia
\and
School of Cyber Science and Engineering, Wuhan University, China
\and 
CSIRO Data61, Australia
\and 
Zark Lab, Singapore
\and
School of Business, Monash University, Malaysia
\and
School of Computer Science, University of Sydney, Australia
}

\maketitle              % typeset the header of the contribution

\begin{abstract}
Agentic AI rivals human capabilities across a wide range of domains.
Looking ahead, it is foreseeable that AI agents will autonomously handle complex workflows and interactions.
Early prototypes of this paradigm are emerging, e.g., OpenClaw and Moltbook, signaling a shift toward Agent-to-Agent (A2A) ecosystems.
However, despite these promising blueprints, critical trust and security challenges remain, particularly in scenarios involving financial transactions. 
Ensuring secure and reliable payment mechanisms between unknown and untrusted agents is crucial to complete a fully functional and trustworthy A2A ecosystem. 
Although blockchain-based infrastructures provide a natural foundation for this setting, via programmable settlement, transparent accounting, and open interoperability, trust and security challenges have not yet been fully addressed.
Hence, for the first time, we systematize blockchain-based A2A payments, e.g., X402, with a four-stage lifecycle: discovery, authorization, execution, and accounting. 
We categorize representative designs at each stage and identify key challenges, including weak intent binding, misuse under valid authorization, payment–service decoupling, and limited accountability. 
We highlight future directions for strengthening cross-stage consistency, enabling behavior-aware control, and supporting compositional payment workflows across agents and systems.

% AI agents are becoming economic actors that transact on behalf of users and organisations, making payment a core capability for emerging \textit{agent-to-agent} (A2A) systems. 
% However, enabling autonomous agents to exchange value introduces fundamental challenges in both trust and security: agents must interact with unknown counterparties without prior trust, while delegated payment authority creates risks of misuse, misalignment, and financial loss. 
% Blockchain-based infrastructures provide a natural foundation for this setting through programmable settlement, transparent accounting, and open interoperability. 
% In this paper, we systematize blockchain-based A2A payments (e.g., X402) through a four-stage lifecycle: discovery, authorization, execution, and accounting. 
% We categorize representative designs at each stage and identify key challenges, including weak intent binding, misuse under valid authorization, payment–service decoupling, and limited accountability. 
% We conclude by highlighting future directions for strengthening cross-stage consistency, enabling behavior-aware control, and supporting compositional payment workflows across agents and systems.


\end{abstract}

% However, once payment authority is delegated to autonomous software, the key challenge is no longer only how to pay, but \textit{how to ensure that the entire payment flow remains correct, secure, and accountable}. 

\section{Introduction}

AI agents are evolving from one-off chatbot interactions into persistent software entities that coordinate across tools, services, and agent-to-agent environments at scale~\cite{wu2024autogen}. 
Emerging agentic systems such as OpenClaw~\cite{openclaw2024,chen2026clawed} and agent-native platforms like Moltbook~\cite{jiang2026humans} illustrate this transition toward ecosystems in which agents interact continuously with external services and other agents. 
More broadly, large language models (LLMs) have enabled agentic systems capable of planning, reasoning, and invoking external tools to carry out tasks over extended periods of time~\cite{islam2026rise,jiang2026sok,valmeekam2023planning,guo2024large,hu2025agentgen,abou2025agentic,plaat2025agentic,yu2026plantwin}. 
These agents act on behalf of human users and organizations to access APIs, acquire data or computational resources, negotiate with other agents, and complete service workflows with limited human intervention~\cite{ray2025review,xu2026agent}. 
As agent operation becomes persistent rather than episodic, payment capabilities become necessary, as coordinating access to resources, services, and other agents inherently requires economic mechanisms~\cite{rothschild2025agentic,conand2025financial}.

However, conventional human-oriented payment infrastructures require payments to be explicitly initiated and authorized by human users, typically through predefined merchant interfaces (e.g., checkout flows or API endpoints)~\cite{hancock1997payment,mastercard_payment_process,w3c_payment_request}. 
This model does not extend naturally to agent ecosystems, where agents autonomously discover services, compose multi-step workflows, and interact with previously unseen counterparties.
Such agent-to-agent interactions instead require payments to be programmable, interoperable across heterogeneous platforms, and verifiable without bilateral trust, while supporting high-frequency, low-value interactions across services and counterparties.
Recent extensions to conventional payment infrastructures, such as Mastercard Agent Pay~\cite{mastercard_agent_pay}, enable AI agents to transact using tokenized payment credentials under delegated authorization constraints. 
However, as they remain tied to card-network authorization and settlement rails, each interaction is processed as an independent authorization, making such systems ill-suited for workloads involving frequent small payments or multi-step tasks due to repeated authorization overhead. 
Hence, to support and enable a fully automated agent-to-agent (A2A) payment system, blockchain-based infrastructures offer a compelling solution. With their programmability, executability, global accessibility, and inherent transaction verifiability, such systems make it possible to embed value transfer directly into automated workflows, eliminating the need for custom bilateral integrations and reducing dependence on fully trusted intermediaries.

% Crucially, they do not maintain a persistent, shared execution context across interactions, limiting their ability to coordinate cross-service workflows and to bind payments to corresponding service outcomes within a unified reference.

We notice that recent blockchain-based systems and design proposals~\cite{vaziry2025towards,birch2025agentic,kite2025agentnative,xu2026agent,xiang2025leveraging,lei2025large} were connecting LLM-enabled agents to on-chain or chain-anchored payment rails, supporting automated settlement, metering, and basic accounting for service interactions. 
But unfortunately, enabling agents to pay autonomously does not by itself ensure that payment intent, authorization, execution, and service outcome remain aligned~\cite{li2026a402,acharya2025secure}. 
Once financial authority is delegated to autonomous agents, errors, misalignment, or adversarial manipulation can lead directly to financial loss, unauthorized spending, and violations of governance constraints~\cite{li2026sok,chen2025chatgpt,deng2025ai,marino2025giving,li2026a402}. 
Recent evidence further shows that agent protocol stacks already expose protocol-logic vulnerabilities and supply-chain style attacks~\cite{lia2asecbench}, while payment-enabled workflows amplify the consequences of prompt injection and interaction manipulation~\cite{debi2026whispers,zhu2025automated}.
At the same time, blockchain payment infrastructures introduce practical constraints related to latency, fees, scalability, and privacy that must still be reconciled with off-chain service execution and provenance tracking~\cite{esmaili2025performance}.

Therefore, in this Systematization of Knowledge (SoK) paper, we, for the first time, systematically examine and evaluate the trust, privacy, and security risks of existing mechanisms for Agent-to-Agent payment systems, particularly in which agents can autonomously or conditionally initiate payments, receive payment-triggered services, or both through blockchain-based infrastructures. 
Particularly, we categorize the identified risks and challenges, including weak intent binding, misuse under valid authorization, payment–service decoupling,and limited accountability, into a four-stage life cycle, spanning discovery, authorization, execution, and accounting.
Our contributions are summarized as follows:

\begin{itemize}
\item We propose a lifecycle model for blockchain-based payments for AI agents, which provides a common abstraction for reasoning about discovery, authorization, execution, and accounting across heterogeneous systems.
\item We organize and systematize the emerging design space by mapping representative mechanisms, system assumptions, and deployment patterns onto this reference model, thereby clarifying how current approaches differ and where their trade-offs arise.
\item We derive a structured view of the risk surface and research gaps in agentic payments, highlighting challenges in delegated spend control, service--payment coupling, accountability, privacy, scalability, and compliance.
\item We point out future directions and possibilities to address the raised issues.
\end{itemize}

\section{Preliminaries}

We distinguish three core abstractions underlying agent-mediated payment systems: large language models (LLMs) as generative components, AI agents as systems that interpret model outputs and execute external actions, and blockchain as the programmable settlement substrate for realizing payment logic~\cite{vaswani2017attention,wei2022chain,yao2023react,schick2023toolformer,nakamoto2008bitcoin,buterin2014ethereum}.

\subsection{LLMs and AI Agents}

LLMs are generative models, typically based on Transformer architectures~\cite{vaswani2017attention}, that produce text and intermediate reasoning traces conditioned on context~\cite{wei2022chain}. 
In this paper, the LLM is treated as a component that generates candidate actions and structured outputs. 
We view AI agents as software systems that orchestrate model outputs, memory, and tool execution to perform multi-step interactions with external services~\cite{yao2023react,schick2023toolformer,park2023generative,qin2024toolllm}. 
A typical interaction loop consists of generating action directives, executing tool calls, and incorporating observations into subsequent steps. 
In the payment setting, this reduces to sequences of tool-mediated interactions through which model outputs are translated into payment-relevant operations.

\begin{figure}[htbp]
    \centering
    \begin{subfigure}[b]{0.44\linewidth}
        \centering
        \includegraphics[width=\linewidth]{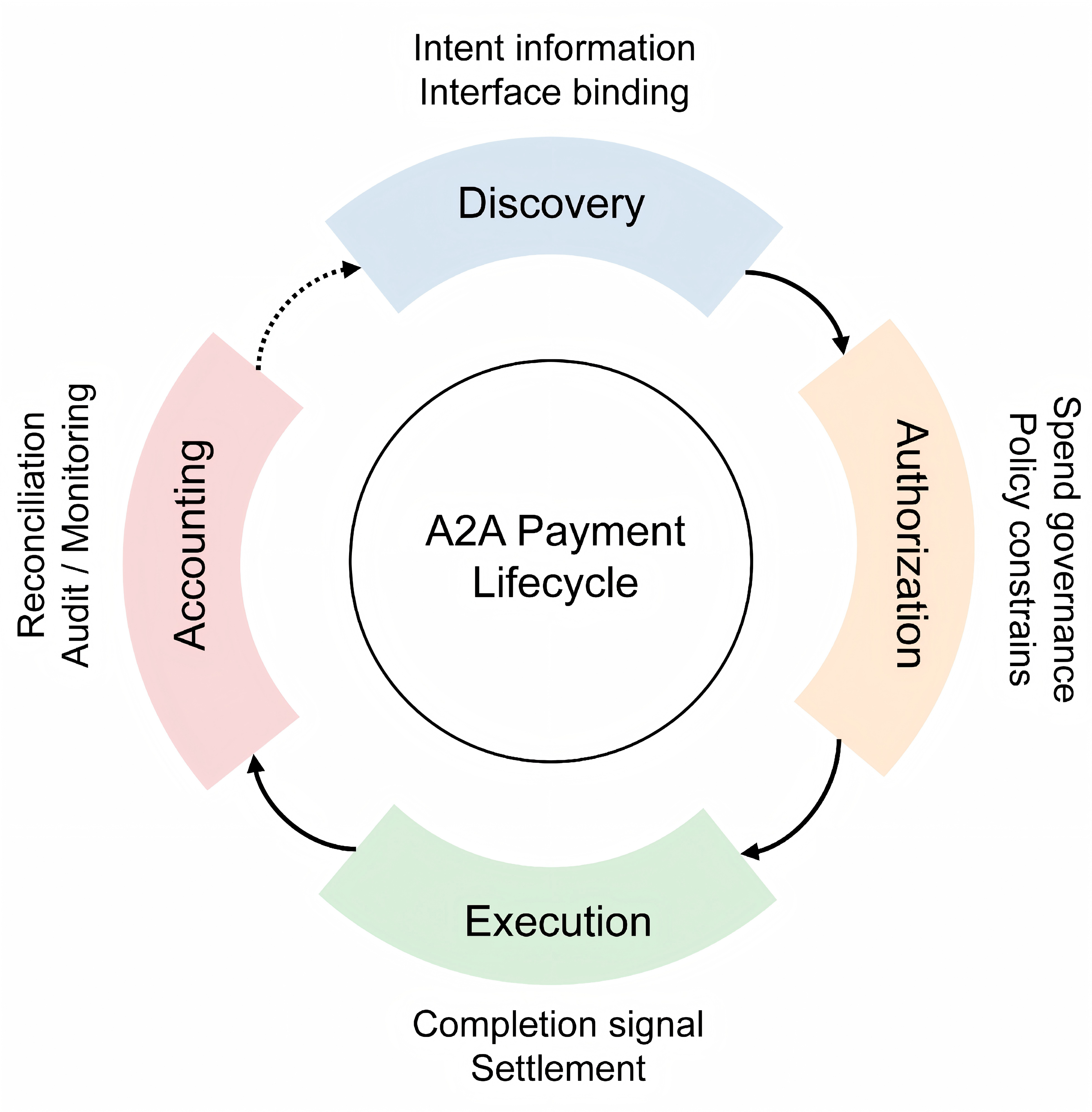}
        \caption{Agentic Payment Lifecycle}
        \label{fig:lifecycle}
    \end{subfigure}
    \hfill
    \begin{subfigure}[b]{0.55\linewidth}
        \centering
        \includegraphics[width=\linewidth]{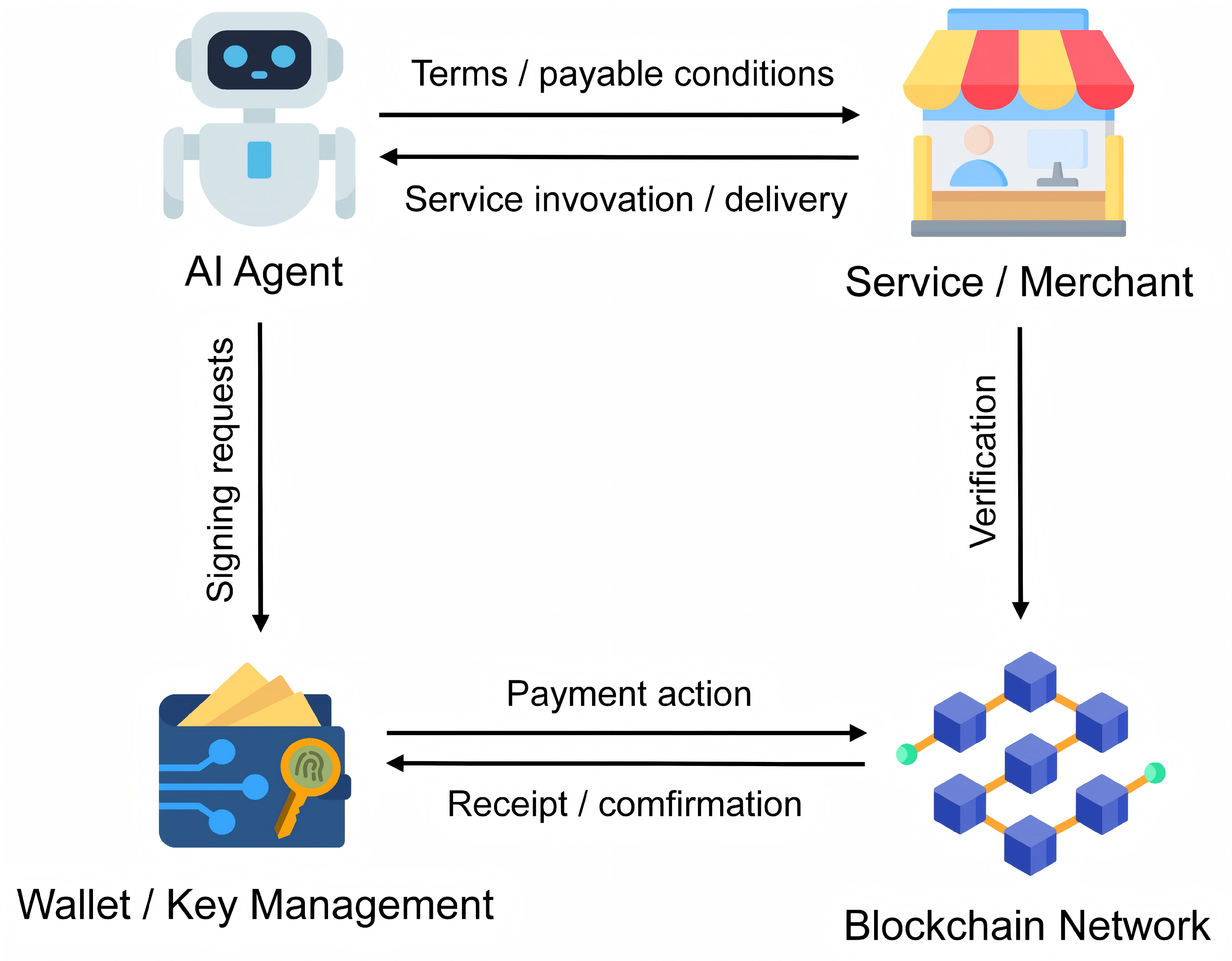}
        \caption{Participant interaction framework}
        \label{fig:framework}
    \end{subfigure}
    \caption{Overview of blockchain A2A payments}
    \label{fig:overview}
    \vspace{-1em}
\end{figure}

\subsection{Blockchain and Programmable Payment}

Blockchains provide a decentralized, append-only settlement layer for transferring value and verifying system state without central intermediaries~\cite{nakamoto2008bitcoin,wang2019sok}. 
Smart-contract platforms such as Ethereum extend this model to programmable transactions, enabling payment conditions, delegation constraints, and execution rules to be enforced directly within the settlement process~\cite{buterin2014ethereum}. 
Recent advances in scaling (e.g., sharding and L2 systems) have reduced latency and transaction costs~\cite{chaliasos2025towards,zhang2023txallo,zhang2025mosaic}, improving the practicality of high-frequency payments. 
Accordingly, we treat blockchain as the infrastructure that enables verifiable and programmable settlement for agent-mediated interactions, while introducing trade-offs in usability, recourse, and risk allocation~\cite{li2026sokstablecoins}.

\section{Agent-to-Agent Payment Lifecycle}

We summarize blockchain-based agent payments using a four-stage lifecycle (Fig.~\ref{fig:lifecycle}), capturing recurring functional stages across existing systems. 
This lifecycle serves as a compact analytical reference for structuring the survey and localizing mechanisms and risks without committing to a specific architecture. 
Complementarily, Fig.~\ref{fig:framework} illustrates the main participants and interactions: payable conditions arise at the agent–merchant interface, payment actions are executed via wallet or key-management components, and settlement occurs on the blockchain, with on-chain receipts supporting confirmation and reconciliation.

\textbf{\ding{182} Discovery.} 
The lifecycle begins with identifying a payable operation and constructing a structured payment intent that binds the payment obligation to its execution context, including the requested resource, applicable terms, and a stable request identifier. 
This representation enables subsequent stages to reference a persistent payment obligation and coordinate execution, retries, and accounting.

\textbf{\ding{183} Authorization.}
The system determines whether an authenticated payment request is admissible under predefined governance rules. 
Delegated spending constraints, such as budget limits, policy controls, and approval requirements, are enforced over the authenticated principal to bound permissible execution. 
The outcome is an authorization decision that governs how the payment intent may be executed.

\textbf{\ding{184} Execution.}
Given authorization, the system realizes the payment intent as a concrete payment action and produces a completion signal that allows the workflow to progress. 
This stage includes handling transaction outcomes such as failures, retries, and settlement delays.

\textbf{\ding{185} Accounting.}
The system maintains records that link payment actions with their originating intents and observed outcomes to support reconciliation, auditability, and operational monitoring. 
This stage provides traceability across the workflow and enables downstream processes such as billing verification and dispute resolution.

\smallskip
The lifecycle separates four fundamental functions, and we use this reference model throughout the SoK to align surveyed approaches and localize risks, limitations, and open challenges to their corresponding stages.

\section{Existing Approaches by Lifecycle Stages}

We analyze existing approaches from a lifecycle perspective. Table~\ref{tab:lifecycle_mapping} maps representative agentic payment systems to the stages they primarily support, motivating a stage-wise decomposition.

\subsection{Discovery and Intent Binding}

The discovery stage determines how agents locate payable counterparties and construct payment intents tied to a specific interaction context.  We organize this stage along two dimensions: (i) the \emph{discovery substrate}, which defines how agents and services are resolved, and (ii) the \emph{intent-binding primitive}, which specifies how payment obligations are associated with the interaction context.

\smallskip
\noindent\textbf{Discovery Substrates (D):} 
Discovery mechanisms differ in how counterparties and service metadata are resolved. 
We distinguish three dimensions: descriptor-based, identity-augmented, and behavior-derived discovery.

\textit{D1. Descriptor-based discovery.}  
Services expose structured capability descriptors (e.g., AgentCards~\cite{agentcard_docs}) that specify supported actions, invocation schemas, and potentially pricing or payment requirements.  
These descriptors may be retrieved directly from service endpoints (e.g., HTTP-accessible metadata) or indirectly via registry-based resolution, where identifiers or descriptor references are anchored on-chain and resolved through smart contracts or indexing services~\cite{vaziry2025towards}.

\textit{D2. Identity-augmented discovery.}
Identity-centric systems enrich discovery with identity and delegation information, such as relationships among users, agents, and sessions. They may also use trust-related signals like claims, stake, or past behavior (e.g., Kite~\cite{kite2025agentnative}, EIP-8004-style mechanisms~\cite{eip8004}).
Skyfire is a concrete example of this approach. Its \emph{Know Your Agent (KYA)} layer and service directory link agent identities and credentials to discoverable services, and bring identity verification into both interaction and payment workflows~\cite{skyfire2025stack}.

\textit{D3. Behavior-derived discovery.}
Discovery can also be driven by behavioral signals observed from prior interactions rather than declared capabilities or semantic descriptions.
In this paradigm~\cite{shi2025sybil}, services are ranked based on interaction traces such as payment flows, where transactions act as implicit endorsements, and reputation is propagated according to the quality of participating agents, transaction value, and temporal recency.

\begin{table}[t]
\centering
\small
\renewcommand{\arraystretch}{1.01}
\setlength{\tabcolsep}{1pt}

\caption{Lifecycle decomposition of existing approaches by stages they primarily support. 
\protect\symP{} primary contribution;  \protect\symS{} discussed; \protect\symN{} not covered.}
\label{tab:lifecycle_mapping}
\vspace{2pt}
\setlength{\tabcolsep}{10pt}
\resizebox{\linewidth}{!}{
\begin{tabular}{c | p{3cm}|cccc}
\toprule
\multicolumn{1}{c}{\textbf{Paradigm}} &
\multicolumn{1}{c}{\textbf{Designs}} & 
\textbf{Disc.} & 
\textbf{Auth.} & 
\textbf{Exec.} & 
\textbf{Acct.} \\
\midrule

\multirow{5}{*}{\makecell[c]{{Service-Specified Payment}}}
& x402~\cite{x402_whitepaper} & \symP & \symS & \symP & \symS \\
& MPP~\cite{stripe_mpp} & \symP & \symS & \symP & \symS \\
& Vaziry \emph{et al.}~\cite{vaziry2025towards} & \symP & \symS & \symP & \symS \\
& Skyfire~\cite{skyfire2025stack} & \symP & \symS & \symP & \symS \\
& Agent Pay~\cite{mastercard_agent_pay} & \symN & \symS & \symP & \symS \\

\cmidrule(lr){1-6}

\multirow{3}{*}{\makecell[c]{{Owner-Specified Payment}}}
& ERC-4337~\cite{erc4337_docs} & \symN & \symP & \symS & \symN \\
& Dimitrov \emph{et al.}~\cite{dimitrov2025leveraging} & \symN & \symP & \symS & \symS \\
& Gorzny \emph{et al.}~\cite{gorzny2025account} & \symN & \symP & \symS & \symS \\

\cmidrule(lr){1-6}

\multirow{2}{*}{\makecell[c]{{Identity-Augmented Payment}}}
& Kite~\cite{kite2025agentnative} & \symS & \symP & \symP & \symS \\
& EIP-8004~\cite{eip8004} & \symP & \symS & \symN & \symN \\

\cmidrule(lr){1-6}

\multirow{3}{*}{Payment Execution Infrastructure
%\makecell[c]{{Payment Execution}\\{Infrastructure}}
}
& Nanopayments~\cite{circle_nanopayments_docs} & \symN & \symS & \symP & \symN \\
& Tempo~\cite{tempo_mainnet} & \symN & \symN & \symP & \symN \\
& Ravi \emph{et al.}~\cite{ravi2026agentic} & \symN & \symS & \symP & \symN \\

\cmidrule(lr){1-6}

\multirow{2}{*}{\makecell[c]{ {Cross-Stage Coupling}\\ }}
& Li \emph{et al.}~\cite{li2026a402} & \symN & \symN & \symP & \symP \\
& Acharya \emph{et al.}~\cite{acharya2025secure} & \symS & \symP & \symS & \symP \\

\bottomrule
\end{tabular}
}
\vspace{-1em}
\end{table}

\smallskip
\noindent\textbf{Intent-Binding Primitives (B):} Intent-binding mechanisms differ in what elements of an interaction a payment obligation is associated with, such as individual service requests or identity and session context.

\textit{B1. Payment-term binding via in-band signaling.}
In x402-style interaction flows~\cite{x402_whitepaper,vaziry2025towards,stripe_mpp}, services respond to requests with payment requirements (e.g., asset type, recipient address, amount) as part of a challenge–response exchange. 
Agents satisfy these requirements by attaching a corresponding payment transaction or proof in subsequent requests, thereby completing the interaction.

\textit{B2. Context and provenance binding.}
Identity-centric designs bind payment authorization to identity and delegation context, such as user identity, agent instance, and session-level constraints (e.g., Kite~\cite{kite2025agentnative}).  
In these systems, payment intents are realized as context-scoped delegation objects whose validity and interpretation depend on the associated identity and session state, enabling actions to be attributed to specific principals and delegated sessions.
More expressive designs make this binding explicit by representing user intent as verifiable objects.  
For example, Acharya~\cite{acharya2025secure} introduces on-chain intent proofs that bind user authorization to payment execution, allowing transactions to be verified against user-approved constraints.

\subsection{Authorization}

Authorization builds on intent binding by enforcing which payment intents may be executed and under what constraints.
It operates over requests that are bound to a principal through authentication (e.g., cryptographic signatures, credentials, or session context), and determines whether the resulting payment intent is admissible under delegated policies.
We organize existing designs along two dimensions: (i) \emph{authorization carriers} (A), which define where and how spending authority is enforced, and (ii) \emph{policy expressiveness} (E), which captures how richly constraints over spending behavior are specified.

\smallskip
\noindent\textbf{Authorization Carriers (A):}
Authorization carriers differ in where spending authority is enforced along the transaction path, such as at contract invocation, asset transfer, or wallet-level validation prior to execution.

\textit{A1. Contract-mediated delegation.}
Agent actions are mediated by smart contracts that act as execution intermediaries and enforce access control over callable functions (e.g., role-based permissions)~\cite{borjigin2025ai}.  
Agents invoke contract functions under predefined roles, and the contract determines whether the invocation is permitted based on its internal logic.

\textit{A2. Allowance- and approval-mediated spending.}
Spending authority is delegated via token allowances or approval mechanisms, where agents are permitted to transfer assets within predefined limits (e.g., ERC-20 approvals).  
Authorization is enforced at the asset layer, where transactions are validated against allowance constraints such as maximum spendable amount or approved spender.  
Authentication is provided by signature-based transaction validation, either through on-chain approval transactions or off-chain signed permits (e.g., EIP-2612~\cite{eip2612}), binding the transfer request to the token holder or an approved spender address.

\textit{A3. Wallet-mediated programmable authorization.}
Account abstraction (AA), as instantiated in ERC-4337~\cite{erc4337_docs,wang2023account}, shifts authorization logic into smart contract wallets that validate user operations prior to execution.  
Each operation is submitted as a \emph{UserOperation} and processed by an entry point contract, which invokes wallet-defined validation logic to check signatures, policies, and contextual constraints before inclusion.    
Such programmable authorization has been explored in both wallet systems and agent-driven designs (e.g.,~\cite{dimitrov2025leveraging,gorzny2025account}), where validation logic is extended to support automation and policy enforcement.

\smallskip
\noindent\textbf{Policy Expressiveness (E):} Policy expressiveness captures what constraints can be specified over delegated spending behavior, ranging from access-level permissions to transaction-level bounds and contextual policies.

\textit{E1. Access-level constraints.}
Policies specify which actions or contract functions an agent is permitted to invoke (e.g., role-based access control)~\cite{borjigin2025ai}.  
Constraints are defined over callable operations, determining whether a given invocation is allowed.  

\textit{E2. Transaction-level constraints.}
These constraints are typically enforced through token allowance or approval mechanisms (e.g., ERC-20 approvals), where transactions are validated against predefined limits at execution time.  More flexible forms of transaction-level constraints can be expressed through signature-based approvals (e.g., EIP-2612~\cite{eip2612} and Permit2~\cite{permit2_docs}), enabling fine-grained control over transaction parameters.

\textit{E3. Contextual and stateful policies.}
Policies incorporate contextual information and state-dependent conditions, such as rate limits, cumulative spending bounds, or delegation scopes, evaluated during validation (e.g., in AA-based systems)~\cite{dimitrov2025leveraging,gorzny2025account}.  
Constraints may depend on interaction context, temporal conditions, or historical state.

\subsection{Execution and Settlement}

The execution stage concerns how payment intents are realized as concrete transactions, how these transactions are submitted and validated, and what conditions define completion for subsequent workflow progression. 
We organize this stage along three dimensions: 
(i) \emph{settlement paths} (S), which determine how and where payments are finalized; 
(ii) \emph{submission and fee orchestration} (O), which determines how transactions are constructed, submitted, and funded; and 
(iii) \emph{access-gating evidence} (G), which determines what observable signals are used to trigger service access or workflow continuation.

\smallskip
\noindent\textbf{Settlement Paths (S):} Settlement paths differ in where payment state is maintained and when settlement is finalized, such as through direct on-chain inclusion or off-chain coordination with deferred settlement.

\textit{S1. Direct on-chain settlement.}
Each interaction is realized as an on-chain transaction, with completion defined by transaction inclusion in the blockchain~\cite{x402_whitepaper,vaziry2025towards,mastercard_payment_process}.  
Agents construct and submit transactions that directly transfer assets to the service provider, and subsequent workflow progression is conditioned on confirmation of inclusion.  
The mechanism operates at the blockchain layer, where the payment state is recorded and finalized on-chain.

\textit{S2. Off-chain–coordinated settlement.}
Payment interactions are executed off-chain through signed updates or bilateral agreements, with final settlement deferred to a later on-chain transaction~\cite{kite2025agentnative,tempo_mainnet}. 
Agents exchange off-chain payment updates that represent incremental transfers, which are later consolidated and settled on-chain as a single transaction.  
The mechanism operates across off-chain coordination and on-chain settlement layers, where the payment state evolves off-chain and is periodically committed on-chain.
% Circle Nanopayments~\cite{circle_nanopayments_docs} exemplifies this design by supporting continuous agent-driven payment updates that are aggregated and settled as batched on-chain transfers.

\smallskip
\noindent\textbf{Submission and Fee Orchestration (O):}
Submission and fee orchestration mechanisms differ in who submits the payment transaction and how execution costs are provisioned. 
We distinguish three patterns: direct client submission, account-abstraction-mediated relaying, and facilitator-mediated submission based on off-chain authorization.

\textit{O1. Client-submitted transactions.}
Agents construct, sign, and submit transactions directly to the network and are responsible for paying transaction fees.  
Execution is tied to on-chain confirmation, with agents managing nonce, gas pricing, and submission timing~\cite{vaziry2025towards}.

\textit{O2. Account-abstraction mediated submission.}
Account abstraction (e.g., ERC-4337 \cite{wang2023account}) introduces an alternative submission flow in which agents produce \emph{UserOperations} relayed by bundlers and executed via an entry point contract~\cite{erc4337_docs,dimitrov2025leveraging,gorzny2025account}. Paymasters may sponsor transaction fees on behalf of agents, decoupling fee payment from the originating account.

\textit{O3. Facilitated submission with off-chain authorization.}
Agents authorize token transfers off-chain using signed transfer approvals (e.g., EIP-2612~\cite{eip2612}, or Permit2~\cite{permit2_docs}), and a facilitator submits the corresponding on-chain transaction on their behalf~\cite{x402_eip2612_gas_sponsoring}. facilitators handle transaction broadcast and may sponsor gas, while recovering costs through the payment flow or application-level fee handling.

\subsubsection{Access-Gating Evidence (G):}

Access-gating mechanisms differ in what observable evidence is required to trigger service execution or workflow continuation, such as on-chain transaction inclusion or off-chain payment state.

\textit{G1. Inclusion-gated access.}
Service access is conditioned on the inclusion of a corresponding on-chain payment transaction~\cite{x402_whitepaper,vaziry2025towards}.  
Agents submit a payment transaction, and services verify its inclusion before proceeding with execution.  
The mechanism operates at the blockchain layer, where access is gated by confirmed on-chain payment records.  
Recent industry protocols such as Stripe's Machine Payments Protocol (MPP) follow a similar model, integrating payment authorization and confirmation into API interaction flows and gating service access on successful payment completion~\cite{stripe_mpp}.  
Such interaction patterns may be deployed over different execution environments, including emerging infrastructures such as Tempo that support crypto-based agent payments~\cite{tempo_mainnet}.

\textit{G2. Off-chain update--gated access.}
Access is granted based on off-chain payment state or deferred-settlement updates, rather than direct on-chain transaction inclusion.  
Services accept intermediate payment evidence as sufficient to continue execution, while final settlement is completed later on-chain.  
Such designs arise in deferred-settlement or liquidity-aware payment protocols~\cite{ravi2026agentic}.

\textit{G3. Proof-based or attestation-gated access.}
Access is granted based on verifiable payment evidence beyond direct transaction inclusion, such as signed receipts, cryptographic proofs, or third-party attestations.  
Agents present verifiable artifacts that attest to payment completion or entitlement, which services validate before execution.  
Such designs are explored in systems that couple payment with verifiable execution or authorization evidence (e.g., A402~\cite{li2026a402}, Acharya~\cite{acharya2025secure}), where access decisions are derived from cryptographically verifiable proofs rather than on-chain inclusion alone.

\subsection{Accounting}

The accounting stage concerns how payments are verified and linked to service outcomes. 
We structure this in two ways: (i) \emph{verification evidence} (V), which defines what constitutes valid evidence of payment, and (ii) \emph{service--payment coupling} (C), which defines how payments are associated with outcomes.

\smallskip
\noindent\textbf{Verification Evidence (V):} Verification evidence differs in what artifacts are used to establish that a payment has occurred, and at which layer such evidence is generated and validated.

\textit{V1. On-chain transaction evidence.}
Blockchain transaction records serve as the primary proof of payment, with completion defined by inclusion in the ledger~\cite{x402_whitepaper,vaziry2025towards,alqithami2026autonomous}.  
Agents or services verify that a transaction transferring the specified asset and amount to a designated recipient has been included on-chain.

\textit{V2. Interaction-level receipts and logs.}
Systems may generate application-level artifacts such as execution logs, interaction traces, or service-issued receipts that record the occurrence of payment-related events.  
Such artifacts are commonly used for auditing and debugging in agent-based systems, where execution traces or policy decision records may be retained for verification or analysis~\cite{alqithami2026autonomous}.

\smallskip
\noindent\textbf{Service-Payment Coupling (C):} 
Service-payment coupling mechanisms differ in the timing and strength of linkage between payment and service execution, ranging from execution-triggered access, to post hoc accountability, and to protocol-level enforcement.

\textit{C1. Execution-trigger coupling.}
Payment confirmation is used as a condition to trigger service execution, establishing a direct dependency between payment events and invocation of service logic~\cite{x402_whitepaper,vaziry2025towards,mastercard_payment_process}.

\textit{C2. Post hoc accountability linkage.}
Mechanisms such as insurance-based accountability layers associate payments with service outcomes after execution through retrospective evaluation and dispute resolution processes~\cite{hu2025insured}.  
These mechanisms support auditing and responsibility attribution by linking payment events to outcomes ex post, rather than enforcing coupling during execution.  

\textit{C3. Enforcement-based coupling.}
Payment settlement and service execution are cryptographically or protocol-level interdependent, such that neither can be completed unilaterally. 
For example, A402~\cite{li2026a402} introduces atomic service channels that bind payment completion to service delivery via adaptor signatures, ensuring that payment is finalized only upon the release of execution-dependent secrets.

\section{A2A Payment Challenges}

We have analyzed risks and limitations above by mapping them into the stages of the agent payment lifecycle, providing a structured view of where vulnerabilities and operational challenges arise across different phases of interaction.  
We then further distinguish technical weaknesses, transaction level mismatches, and broader governance constraints in each stage and discuss separately on agent authentication, including identity-related issues, and future directions.
\subsection{Discovery and Intent Binding Risks}
\label{sec:discovery}

Discovery and intent binding use exposed metadata to produce payment intents. However, neither of them guarantees correct execution or institutional admissibility.

\smallskip
\noindent\textbf{Technical level.}
Discovery via endpoints or registries (D1) constructs payment intents from externally exposed metadata without authenticating semantic identity~\cite{ray2025review,muttoni2025agent}. 
This creates a phishing-like surface where adversaries expose valid-looking descriptors through counterfeit endpoints, typosquatted domains, spoofed Agent Cards, or deceptive registry entries~\cite{narajala2025toolsquatting,ferrag2025protocolexploits,lin2025baid,louck2025protocolsecurity}. 
As a result, agents may form syntactically valid payment intents for malicious or nonexistent services, binding payments to the wrong counterparty.

\smallskip
\noindent\textbf{Transaction level.}
Intent-binding mechanisms, for example, in-band payment signaling (B1), embed payment conditions in the request. 
However, they do not reliably tie payments to a specific session, a concrete outcome, or verifiable fulfillment. Acharya~\cite{acharya2025secure} addresses this issue by proposing a stronger design based on decentralized identity, on-chain intent proofs, and attested execution, but it remains a proposal and has not been deployed in mainstream systems.
As a result, a payment may still be valid even when the interaction is replayed, mismatched, or produces missing, partial, or incorrect service outcomes.

\smallskip
\noindent\textbf{Legal and institutional level.}
Discovery selects counterparties based on reachability and functional compatibility. 
However, it neither checks regulation requirements nor institutional admissibility, such as jurisdiction, sanctions, licensing, or platform policies. 
Such checks are rarely performed before intent formation, which prevents service resolution from governance constraints emphasized in agent economies and interoperable systems ~\cite{hadfield2025economy,chaffer2025can,hu2025inter}. 
As a result, agents may execute technically valid payments to counterparties, but they are institutionally prohibited.

\begin{tcolorbox}[colback=gray!10, colframe=black!40, boxrule=0.5pt, arc=2pt]
\textbf{Insight.} Discovery anchors payment to externally described metadata rather than execution semantics or admissibility. 
It may lead to errors happening at this stage, which would propagate downstream, such that later stages may consistently execute and record payments that are internally valid yet fundamentally misaligned with the intended interaction.
\end{tcolorbox}

\subsection{Delegated Authorization and Spend Control Risks}
\label{sec:authorization}

Authorization enforces transaction-level admissibility under delegated control, but does not capture the broader behavioral meaning of spending over time.

\smallskip
\noindent\textbf{Technical level.}
Authorization validates transactions at submission but assumes that transaction generation is trustworthy. 
If the agent or its environment is compromised through prompt injection~\cite{liu2024formalizing}, model manipulation~\cite{zhang2024instruction}, key leakage~\cite{lan2026silent}, software vulnerabilities~\cite{10.1145/3658644.3690338}, or social engineering~\cite{greshake2023not}, adversaries may generate transactions that satisfy authorization policies. 
Because delegated authority is encoded as persistent rules with reactive revocation, policies that are initially valid may become unsafe in operation, but resulting transactions remain admissible.

\smallskip
\noindent\textbf{Transaction level.}
Authorization policies (E1--E3) constrain individual transactions (e.g., amount, recipient, rules)~\cite{x402_whitepaper,gorzny2025account}. However, they do not capture the execution history, cumulative spend, or multi-step strategies. 
Therefore, sequences of valid transactions may violate intended spending boundaries through repetition, fragmentation, or timing manipulation, and this effect is further amplified by strategic agent behavior~\cite{long2025emodebt}. 
This limitation may extend to the multi-entity setting. Considering multiple colluding agents, they distribute actions across identities while remaining locally compliant, and reputation or stake mechanisms provide signals but do not enforce correctness of composed interactions~\cite{joseph2025agency}.

\smallskip
\noindent\textbf{Legal and institutional level.}
The delegated authorization does not ensure continued alignment with user consent or institutional expectations. 
Standing delegation is typically granted ex ante. 
However, the execution of subsequent payments lacks re-evaluation of purpose or cumulative impact. It creates a gap between formal permission and substantive consent. 
Thus, transactions may remain authorized, but actually exceed what users intended or would approve under evolving conditions.

\begin{tcolorbox}[colback=gray!10, colframe=black!40, boxrule=0.5pt, arc=2pt]
\textbf{Insight.} Authorization acts as a local validity filter over individual transactions rather than a mechanism for preserving behavioral correctness over time, and cannot enforce consistency across sequences, delegation contexts, or institutional expectations. 
Thus, formally valid authorization may continue to legitimize spending even after its underlying trust assumptions have broken down.
\end{tcolorbox}

\subsection{Execution and Settlement Risks}
\label{sec:execution}

Execution turns authorized intents into payment transactions. 
However, it treats payment success as completion, rather than the service completion in the real world. Once the payment is settled, the system moves forward even if the service is missing. 
In this sense, execution confirms that payment has been completed, without checking the promised result.

\smallskip
\noindent\textbf{Technical level.}
Execution in blockchain systems is not instantaneous or deterministic, since transaction inclusion is subject to latency, reordering, and probabilistic confirmation~\cite{esmaili2025performance}. 
Agents reacting to intermediate signals may observe delayed or inconsistent states, leading to retries or incorrect workflow progression. 
Execution also depends on auxiliary infrastructure (e.g., bundlers, relayers, paymasters)~\cite{erc4337_docs,dimitrov2025leveraging}. 
It extends the trust boundary beyond the base protocol, where failures or manipulation may degrade reliability even if the underlying chain is secure.

\smallskip
\noindent\textbf{Transaction level.}
The settlement of payment does not guarantee completion. 
There exist some off-chain architectures (S2) that improve throughput by deferring global settlement~\cite{kite2025agentnative}. 
However, they introduce asynchronous, locally visible states that cannot be globally synchronized in time. 
Consequently, parties are left to interpret inconsistent signals, such as off-chain state updates versus on-chain inclusion. Thus, it is difficult to establish a single, authoritative finality point for transactions.

More fundamentally, payment completion is weakly coupled to service completion, since the execution only verifies a successful transfer, not correct delivery. 
In inclusion-gated models such as x402~\cite{x402_whitepaper}, completion is defined by transaction inclusion. 
In off-chain models, it is defined by the acceptance of intermediate states. 
In both cases, financial finality does not establish fulfillment, allowing providers to receive valid payment even if they do not deliver correct outcomes. 
There exist some mechanisms, such as A402~\cite{li2026a402} and liquidity-aware or deferred-settlement protocols~\cite{ravi2026agentic}. They aim for tighter coupling but are still confined to execution-layer coordination. 
Furthermore, they do not eliminate inconsistent interpretations of completion across workflows.

\smallskip
\noindent\textbf{Legal and institutional level.}
In the sense of governance, execution does not determine the payment completion. 
Since settlement signals diverge from service outcomes, there is no shared anchor for different parties to assess fulfillment or liability. 
This gap would be amplified in agentic settings, because payment may be triggered automatically even if the service has not been delivered. 
Thus, execution can produce a valid settlement event but leaves economic and institutional completion unresolved.

\begin{tcolorbox}[colback=gray!10, colframe=black!40, boxrule=0.5pt, arc=2pt]
\textbf{Insight.} Execution provides payment-completion signals rather than a shared notion of workflow completion, and does not ensure consistent observation, agreement on finality, or correct service delivery. 
Thus, payment finality can be operationally valid yet semantically incomplete.
\end{tcolorbox}

\subsection{Accountability and Privacy Risks}
\label{sec:accounting}

Accounting records how payments, outcomes, and responsible actors relate to each other, but the linkage is often incomplete and hard to verify.

\smallskip
\noindent\textbf{Technical level.}
Blockchain serves as an immutable ledger to record transactions, but it cannot capture the off-chain execution details or decision processes. 
Transactions are tied to cryptographic identities, yet the causal chain, including interactions, external inputs, model reasoning and planning, is not recorded in a verifiable form~\cite{guo2024large,karim2025ai}. 
Thus, accounting establishs that a transfer occurred, but not why it occurred or which component in the user--agent--service chain was responsible, leaving causal attribution under-specified.

\smallskip
\noindent\textbf{Transaction level.}
Accounting should evaluate whether payments correspond to successful service fulfillment. However, current systems only loosely link on-chain payment evidence (V1) with off-chain outcomes via logs or receipts. 
For example, x402-style flows prove payment occurrence~\cite{x402_whitepaper}, but their service outcomes are recorded separately. 
It only yields a \textit{post hoc} correlation rather than unified binding. 
Consequently, valid payments may lack verifiable evidence of successful service delivery, leading to ambiguity in audit and dispute resolution. 
The core issue is the absence of an effective binding mechanism that can make payment and outcome jointly accountable.

\smallskip
\noindent\textbf{Privacy and institutional level.}
Records that support accountability may also reveal sensitive personal information.
Transparent transaction histories expose interaction patterns, counterparties, and timing, which may allow others to infer workflows or decision strategies~\cite{bhutta2021survey,deng2025ai}.
This creates a basic trade-off. Greater transparency improves verification and auditing, but causes information leakage. Stronger privacy reduces exposure, but provides less evidence for accountability.

\begin{tcolorbox}[colback=gray!10, colframe=black!40, boxrule=0.5pt, arc=2pt]
\textbf{Insight.} 
Accounting provides reliable payment records. 
But it does not guarantee the linkage among payment, outcome, and responsibility. 
Since evidence (V), coupling (C), and attribution remain only partially aligned, existing systems can only reconstruct causality post hoc, leading to a trade-off between incomplete attribution and excessive exposure. 
Thus, verifiable outcomes and records cannot balance accountability and privacy protection.
\end{tcolorbox}

\subsection{Authentication and Identity Risks}

In existing agentic payments, authentication verifies access but does not establish identity or attribution. 
NIST distinguishes identity proofing, authentication, and federation as separate functions~\cite{nist_sp800_63_4}, a distinction that becomes more critical in agentic settings with delegated authority and cross-domain interactions~\cite{south2025identity}. 
Thus, a system may authenticate a key, account, or endpoint, yet remain blind to whether that action originates from the intended agent, aligns with a correct principal, or represents a stable identity across workflows.

In principle, meaningful identity should be able to handle the fields from protocol authentication to institution compliance. 
For example, FATF requires virtual-asset activity to support AML/CFT controls. So the counterparty attribution and screening are necessary~\cite{fatf_va_vasp_2024}. 
Other systems, such as Coinbase KYT and Tracer, perform transaction screening, risk scoring, and entity attribution through address analysis and fund-flow tracing~\cite{coinbase_compliance_2022}. 
However, even when interactions are authenticated, identity may still be too weakly attributed to support admissibility or accountability.

\begin{tcolorbox}[colback=gray!10, colframe=black!40, boxrule=0.5pt, arc=2pt]
\noindent\textbf{Insight.} Authentication verifies access but does not establish attribution, leaving systems unable to determine the responsible principal or assign accountability for authenticated actions.
\end{tcolorbox}

\subsection{Future Directions}

Based on the above challenges, we identify a structural gap that agentic payment systems do not maintain consistency, control, and security across lifecycle stages. It points to three complementary directions.

\smallskip
\noindent\textbf{Consistency: ensuring that payments correspond to actual outcomes.}
In agentic settings, consistency needs protocol-level support through a stronger \emph{payment--service binding}.
One promising direction is to use a shared append-only execution record that persists across the full lifecycle. 
It achieves commitment of intent during discovery, adds policy decisions to the authorization, attaches settlement references at execution, and records the outcome evidence when accounting.
As a result, all stages refer to the same anchored record. 
It enables the end-to-end consistency checking via a single execution trace.

\smallskip
\noindent\textbf{Control: governing behavior, not just individual transactions.}
Future systems should not judge each payment in isolation. Instead, they should take past behavior into account, such as how much an agent has spent, how often it interacts, and with whom it interacts. They should also be able to tighten restrictions or revoke permissions when behavior starts to change. Incorporating a decentralized identity management system into it is a promising solution direction.

\smallskip
\noindent\textbf{Composition: coordinating payments across workflows.}
Agentic payment can be regarded as a composition of interdependent actions across agents, systems, and execution contexts. 
For example, in multi-hop workflows, delegated tasks propagate payment obligations. 
It requires the dependency-aware consistency across many chained and concurrent interactions. 
This also spans heterogeneous execution environments, where different parts of a workflow execute across mixed settlement rails, such as on-chain settlement and off-chain service provisioning. It makes execution and accounting inherently cross-system. 

\smallskip
\noindent\textbf{Formation: negotiating payable terms across interactions.}
We observe that payment terms may not always be fixed upfront. 
For example, agent-mediated interactions can involve iterative or multi-round negotiation over price, volume, or service scope, as reflected in evolving x402-style pricing models such as ``up-to'' pricing and negotiated schemes~\cite{x402_faq,x402_negotiated_pr}. 
Therefore, lifecycle models should account for how payable conditions are formed across interactions, rather than assuming they are fully specified prior to authorization and execution.

\section{Conclusion}
In summary, while blockchain enables agents to make payments, it does not inherently guarantee their correctness. We systematize this space through a four-stage lifecycle decomposition model and show that existing designs provide partial guarantees at individual stages, failing to preserve correctness across the end-to-end workflow. The key gaps lie in intent binding, delegated control over evolving agent behavior, and the accountable linkage between payment and service outcomes. By identifying these limitations and outlining future research directions, we aim to contribute to the development of a secure and reliable A2A payment ecosystem.

\section*{Acknowledgment}
This research is supported by the National Research Foundation, Singapore and Infocomm Media Development Authority under its Trust Tech Funding Initiative. Any opinions, findings and conclusions or recommendations expressed in this material are those of the author(s) and do not reflect the views of National Research Foundation, Singapore and Infocomm Media Development Authority.

\bibliographystyle{unsrt}
\bibliography{ref}

\end{document}